\documentclass[%
 aip,
 amsmath,amssymb,
 reprint,%
]{revtex4-1}

\usepackage{graphicx}
\usepackage{dcolumn}
\usepackage{bm}
\usepackage[utf8]{inputenc}
\usepackage[T1]{fontenc}
\usepackage{mathptmx}
\usepackage{etoolbox}
\usepackage{xcolor}

\makeatletter
\def\@email#1#2{%
 \endgroup
 \patchcmd{\titleblock@produce}
  {\frontmatter@RRAPformat}
  {\frontmatter@RRAPformat{\produce@RRAP{*#1\href{mailto:#2}{#2}}}\frontmatter@RRAPformat}
  {}{}
}%
\makeatother

\begin{document}


\title[Collimation of diamagnetic laser-driven plasma outflows by an ambient magnetic-pressure gradient]{Collimation of diamagnetic laser-driven plasma outflows by an ambient magnetic-pressure gradient}

\author{Yigeng Tian}
\author{Chung Hei Leung}
\author{Arijit Bose*}
\author{Riddhi Bandyopadhyay}
\author{Michael A. Shay}
\author{William H. Matthaeus}
\email{bose@udel.edu}
\affiliation{Department of Physics and Astronomy, University of Delaware, Newark, Delaware 19716, USA}



\date{\today}

\begin{abstract}
We present magnetohydrodynamic simulations of laser-driven plasma outflows propagating along an externally applied poloidal magnetic field, designed to mimic coronal open-field plasma jets. Using the FLASH code with non-ideal terms (resistivity, Biermann battery, and Nernst advection) included, we model a CH target driven by a 3$\omega$ (351 nm) beam delivering 5 kJ over 10 ns and a uniform background field $\text{B}_0$ = 0–50 T. Under these conditions, the expanding plume develops a central low-density diamagnetic cavity bounded by a high-magnetic-pressure shell. Magnetic flux is advected from the plume center to its edge, and azimuthal diamagnetic currents form that decrease fields inside the cavity and amplify fields outside, producing a radial magnetic-pressure gradient that exerts an inward $\text{J}\times \text{B}$ force and radially confines the flow. \textcolor{black}{We show that the collimation strengthens with increasing applied magnetic field, as stronger fields reduce the plasma $\beta$ and correspondingly enhance the confining 
$\text{J}\times \text{B}$ force.} 
We discuss scaling to solar coronal jets and argue that low-$\beta$, magnetic-pressure-dominated ambient conditions promote similar diamagnetic cavity formation and $\text{J}\times \text{B}$ collimation in coronal outflows.
\end{abstract}


\maketitle

\section{Introduction}\label{section1}
\textcolor{black}{The inner solar corona is a low-$\beta$, sub-Alfvénic plasma regime in which magnetic forces dominate the dynamics, producing a highly structured magnetic environment.} Within this region, collimated outflows such as jets and streamers are frequently observed along open magnetic field lines and are linked to acceleration of energetic particles, coronal heating, and solar wind acceleration\cite{Raouafi+2016,Shen+2021,Joshi+2020}. \textcolor{black}{Images} of the solar corona reveal that the density structure in the corona transitions from anisotropic "striae", generated by the variability of density across magnetic flux tubes, to more isotropic “flocculae” at distances of several tens of solar radii\cite{DeForest+2016}. This textural shift occurs \textcolor{black}{in the vicinity of $\beta \approx 1$ regions, with the} solar wind \textcolor{black}{transitioning from a $\beta \leq 1$} into a $\beta \geq 1$ regime, permitting isotropic expansion \textcolor{black}{at larger radii \cite{Chhiber+2018}.}

Furthermore, \textcolor{black}{the} coronal magnetic flux tubes exhibit diamagnetic behavior, wherein surface diamagnetic currents suppress the internal field while enhancing the external field\cite{Zaitsev+2005,Parkhomov+2018}. This supports the hypothesis that the lack of lateral expansion in solar jets \textcolor{black}{in the lower corona} is \textcolor{black}{possibly because the jet plasma pressure is smaller than the magnetic} pressure of the surrounding open field lines\cite{Moore+2013}. \textcolor{black}{However it is important to note, that while in-situ space missions like NASA’s Parker Solar Probe (PSP) are providing unprecedented data on plasma properties in the solar corona, they do not capture the global morphology and time evolution of the coronal outflows and accompanying magnetic field structure. These large-scale features are essential for understanding the transition in structure and fluctuations in the corona, which have been hypothesized as a potential pathway for the onset of turbulence and its connection to coronal heating \cite{DeForest+2016, Chhiber+2018}.} This work uses laser-plasma simulations to \textcolor{black}{model a laboratory platform for 'controlled' experiments that allow us to potentially} isolate how diamagnetism and magnetic pressure gradients contribute to radial confinement of an expanding plume — a configuration intended to emulate a coronal outflow that follows open field lines.

\begin{figure*}
\includegraphics[width=16cm]{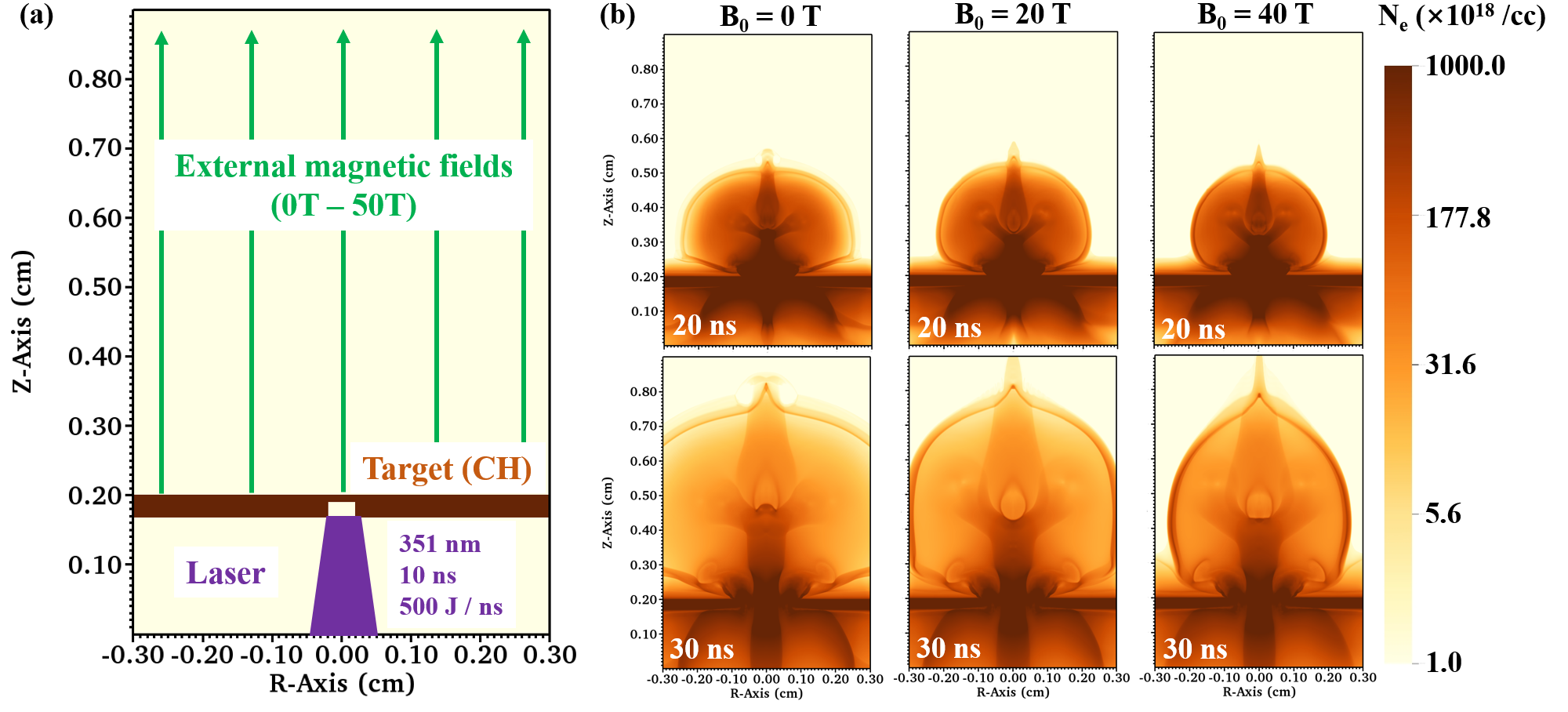}
\caption{(a) Initial simulation setup. A polystyrene (CH) target, comprising a 100 $\mu$m foil and a 230 $\mu$m washer with a 400 $\mu$m central hole, is driven by a laser beam (purple). A uniform poloidal magnetic field (green arrows) is applied. (b) Electron number density (log scale) for outflows at different applied magnetic field strengths. By 30 ns, outflows with $\text{B}_0$ = 0 and 20 T have expanded beyond the computational domain radially.}
\label{fig:1}
\end{figure*}

\textcolor{black}{Previous laboratory experiments have explored magnetically mediated collimation of plasma outflows across a range of platforms, investigating the roles of magnetic pressure gradients and $\mathbf{J}\times\mathbf{B}$ forces. In pulsed-power systems, astrophysically relevant jets formed in Z-pinch configurations are collimated through} compression \textcolor{black}{by} surrounding toroidal magnetic fields \cite{Lebedev+2005,Kalashnikov+2021}. Ambient magnetic pressure \textcolor{black}{has also been shown to enhance} the collimation of rotating plasma jets \cite{Valenzuela-Villaseca+2024}, \textcolor{black}{while plasma gun experiments demonstrate that increasing the applied magnetic field strength extends the stable collimation length prior to the onset of kink instabilities \cite{Brady+2012}.}

\textcolor{black}{In laser-plasma experiments, ablation-driven outflows produced on the laser-irradiated side of the target, in the presence of an ambient poloidal magnetic field, form collimated magnetic-nozzle jets, as demonstrated in studies motivated by young stellar object (YSO) jets \cite{Ciardi+2013,Albertazzi+2014,Higginson+2017,Lei+2020,Revet+2021}. These flows exhibit collimation in parallel field \cite{Ivanov+2019}}, divergent fields \cite{Zemskov+2024}, \textcolor{black}{redirection under} field misalignment \cite{Revet+2021}, and knot-like structures \cite{Lei+2020}, \textcolor{black}{characteristic of nozzle-driven dynamics. Complementary experiments in perpendicular-field configurations have demonstrated diamagnetic cavity formation mediated by flux expulsion, particularly in studies of magnetized Rayleigh–Taylor instability \cite{Zemskov+2025,Yao+2022} and collisionless shocks \cite{Schaeffer+2017,Bondarenko+2017b,Behera+2022,Niemann+2013,Dorst+2022,Bondarenko+2017,Schaeffer+2022,Rovige+2024,Cruz+2023}. These systems typically involve weakly collisional plasmas and are interpreted within kinetic frameworks, in contrast to the magnetohydrodynamic (MHD) regime relevant to coronal outflows.}

\textcolor{black}{Despite these advances, existing laboratory studies predominantly probe plasma produced on the laser-irradiated side of the target, where high temperatures and velocities lead to $\beta \gtrsim 1$, limiting magnetic confinement and resulting in short-lived diamagnetic structures and extended nozzle jets analogous to YSO outflows. In contrast, coronal outflows evolve under low-$\beta$ conditions and propagate along open magnetic field lines, where magnetic pressure governs the dynamics. This regime is more closely analogous to the diamagnetic cavity itself, rather than the downstream magnetic-nozzle jet. Here, we isolate this regime by studying low-$\beta$ plasma expansion driven by rear-side laser irradiation into an ambient magnetic field, where diamagnetically amplified magnetic pressure governs collimation.}

\textcolor{black}{Only a limited number of studies have examined rear-side plasma expansion into an ambient magnetic field \cite{Mabey+2019}, demonstrating partial collimation; however, the roles of plasma diamagnetism and magnetic-pressure–driven confinement remain unquantified. Conflicting observations—such as the disruption of collimated outflows in relatively weak applied fields\cite{Manuel+2015,manuel+2019}—further indicate that additional MHD processes may govern the dynamics in this regime.}

\textcolor{black}{Therefore, a key gap remains: collimation mediated by diamagnetically amplified ambient magnetic pressure has not been directly investigated in a scaled coronal outflow configuration, where a low-$\beta$, expanding plasma plume evolves along an externally imposed magnetic field. The formation, evolution, and dynamical role of the diamagnetic cavity in regulating collimation—and its transition to deflection under non-uniform magnetic topology—remain unresolved. Here, we address this gap by isolating the role of diamagnetically amplified magnetic pressure in governing collimation in the low-$\beta$ rear-side expansion regime.}

\begin{figure}
\includegraphics[width=8cm]{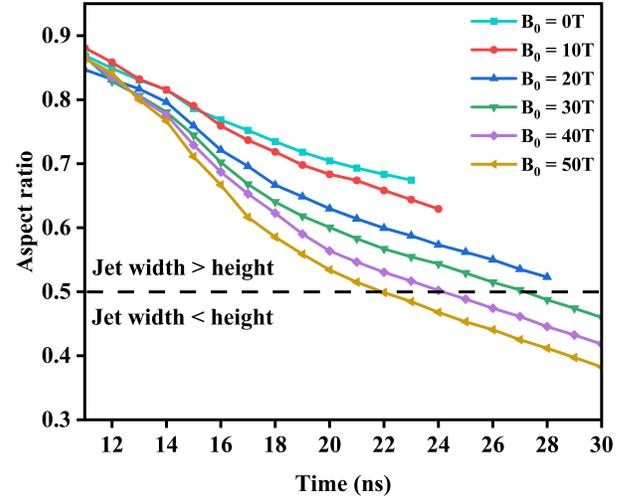}
\caption{Time evolution of the plasma jet aspect ratio (10–30 ns) for applied magnetic fields ($\text{B}_0$) from 0 to 50 T. A lower aspect ratio indicates stronger collimation. The data for $\text{B}_0$ = 0, 10, and 20 T terminate at 23, 24, and 30 ns, respectively, when the jet's radial expansion exceeds the computational domain, making the width unmeasurable.}
\label{fig:2}
\end{figure}

Given the vast \textcolor{black}{difference in spatial and temporal scales between the laboratory experiments and the coronal outflows,} scaling is essential to ensure the \textcolor{black}{validity of the key MHD physics in the} laboratory studies\cite{Drake+2018}. The widely adopted "Ryutov scaling"\cite{Ryutov+2000,Ryutov+2001} preserves \textcolor{black}{MHD equations and} key dimensionless parameters \textcolor{black}{similarity}, such as the plasma beta and Alfvén velocity. When scaling coronal conditions to the laboratory, coronal plasmas are characterized by low densities ($10^9 \text{cm}^{-3}$) and weak magnetic fields (a few Gauss)\cite{Yang+2020}, whereas laser-produced plasmas have much higher densities ($10^{19}-10^{20} \text{cm}^{-3}$). To maintain consistent Alfvén velocities, laboratory \textcolor{black}{studies} must therefore employ high magnetic fields ($10^5$ Gauss).

\begin{figure*}
\includegraphics[width=16cm]{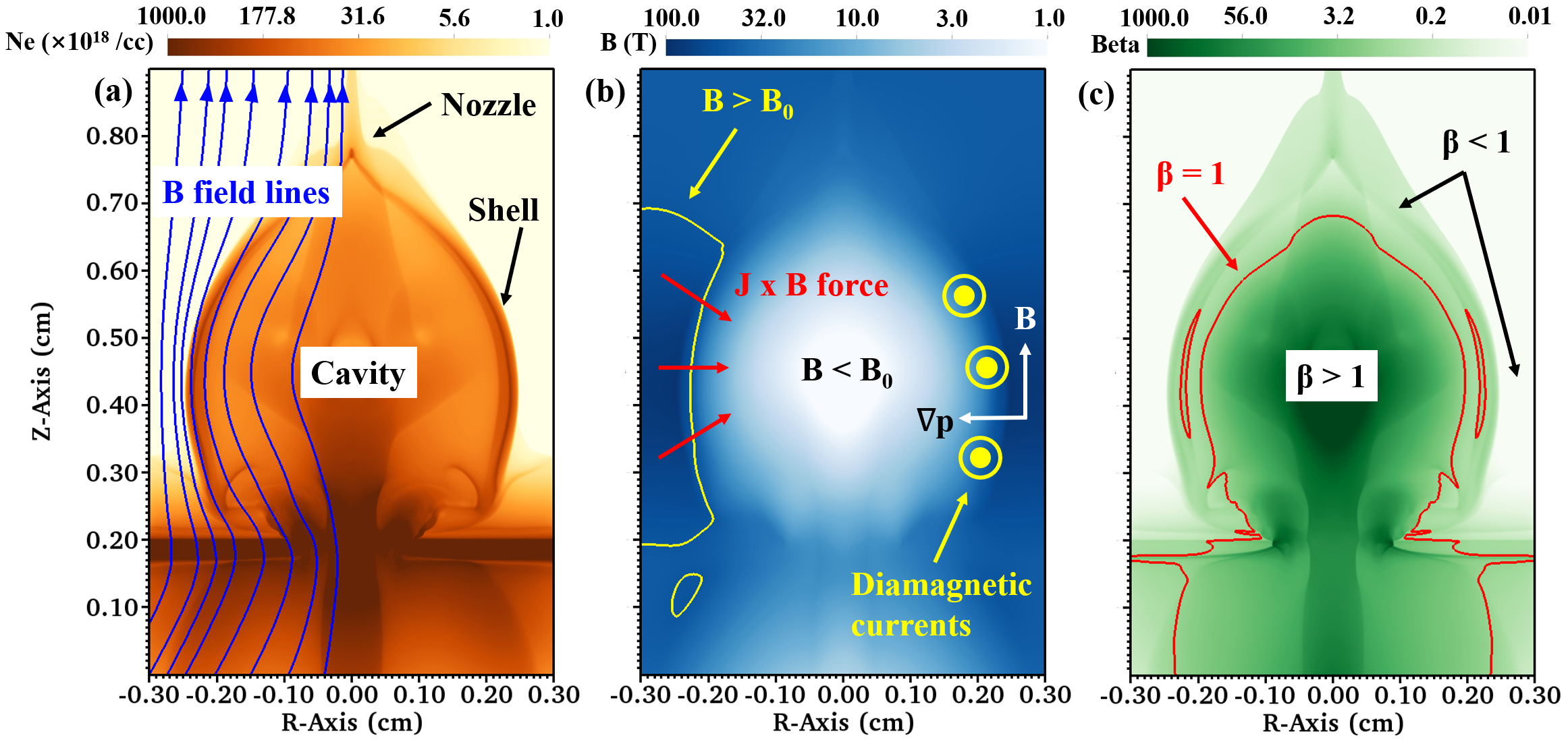}
\caption{For the $\text{B}_0$ = 50 T case at 30 ns: (a) Logarithm of the electron number density, annotated to show the nozzle jet, high-density shell, diamagnetic cavity, and magnetic field lines. (b) Logarithm of the magnetic field strength, with a schematic overlay indicating the high magnetic pressure region ($\text{B} > \text{B}_0$), the confining $\text{J}\times \text{B}$ force, and the diamagnetic current. (c) Logarithm of the plasma $\beta$, with the red contour line marking $\beta$ = 1.}
\label{fig:3}
\end{figure*}

This work \textcolor{black}{is aimed to establish a controlled laboratory platform to investigate} the collimation mechanism of plasma outflows in applied poloidal magnetic fields, \textcolor{black}{by} simulations designed for the OMEGA laser facility\cite{Boehly+1997}. Section~\ref{section2} describes the simulation code and initial configuration. Section~\ref{section3} presents the morphology of the collimated outflow in the external magnetic field. The collimation mechanism and the evolution of the magnetic field are analyzed in Section~\ref{section4}, followed by a discussion of the astrophysical relevance of the simulations in Section~\ref{section5}.

\section{Simulation setup}\label{section2}
The simulations were performed using the FLASH code (version 4.7.1) \cite{Fryxell+2000}. The initial setup, illustrated in Fig.~\ref{fig:1}(a), features a polystyrene (CH) target comprising a 100 $\mu$m-thick foil and a 230 $\mu$m-thick washer, each 6 mm in diameter. A central hole (400 $\mu$m diameter) in the washer confines the laser energy to a specific area on the foil. The system, initialized at room temperature and filled with tenuous helium gas (density: $1.0\times 10^{-6} \text{g/cm}^3$), is driven by a vertical 3$\omega$ (351 nm) \textcolor{black}{super-Gaussian} laser beam \textcolor{black}{(400$\mu$m spot radius)} delivering 5 kJ over 10 ns, \textcolor{black}{where the laser pulse is in square shape.} Externally applied, uniform poloidal magnetic fields ($\text{B}_0$ = 0–50 T) emulate open coronal field lines. Although the simulation setup is for the OMEGA laser facility, this design is versatile and can be easily adapted for use in various laser or pulsed power facilities, provided the appropriate scaling in plasma conditions is applied. In FLASH simulations, the 2D-cylindrical domain (0.3 cm × 0.9 cm in R-Z) is evolved with adaptive mesh refinement (AMR), achieving a peak resolution of 11.7 $\mu$m. \textcolor{black}{The simulations incorporate separate ion, electron, and radiation temperatures, and non-ideal MHD effects\cite{Tzeferacos+2015} such as resistive diffusion\cite{Braginskii+1965}, Biermann battery, and Nernst\cite{Davies+2021}.} The simulations employ the unsplit staggered mesh (USM) MHD solver\cite{Lee+2013}, the HLLD Riemann solver\cite{Miyoshi+2005}, and multigroup radiation diffusion with six \textcolor{black}{energy groups} from 0.1 eV – 100 keV.

\section{Morphology of plasma outflows in the presence of external magnetic fields}\label{section3}
The morphology of plasma outflows under various external poloidal magnetic fields ($\text{B}_0$) is illustrated by the electron number density ($\text{N}_\text{e}$) in Fig.~\ref{fig:1}(b). During the laser heating phase (0–10 ns), \textcolor{black}{the laser irradiates the target from below. The ablated plasma (below the laser critical density) rapidly expands and exits the computational domain on the laser-incident side. Coincident with this, a supersonic jet is generated on the target's rear-side and subsequently propagates upward into the imposed background magnetic field.} While the axial propagation speed remains similar across different $\text{B}_0$ strengths, the outflow width decreases markedly with increasing $\text{B}_0$, indicating that radial expansion is suppressed by the magnetic field. This results in the collimation of the outflow into a narrow jet, an effect that strengthens with the applied field intensity. As shown in Fig.~\ref{fig:3}(a), the outflow structure comprises a low-density cavity bounded by a high-density shell, the low-density protrusion above the cavity is a magnetic nozzle jet that is not the focus of this study, those \textcolor{black}{cavity} features are consistent with other \textcolor{black}{previous studies}\cite{Ciardi+2013,Albertazzi+2014,Revet+2021,Lei+2020,Malko+2024}. 

\textcolor{black}{The collimation effect is quantified by} using the outflow's aspect ratio, defined as the jet radius at half-height divided by the jet height (a lower ratio indicates stronger collimation). Fig.~\ref{fig:2} shows that differences in aspect ratio become significant after 16 ns. The ratio decreases over time and with higher $B_0$, confirming the qualitative observations from Fig.~\ref{fig:1}(b). A key transition occurs when the aspect ratio falls below 0.5, indicating that the poloidal expansion exceeds the radial expansion. This transition is achieved earlier for stronger fields: at 27, 24, and 22 ns for $\text{B}_0$ = 30, 40, and 50 T, respectively, quantitatively demonstrating that collimation is enhanced by stronger magnetic fields.

\section{Collimation mechanism}\label{section4}
Although laser-produced plasma outflows are observed to collimate under applied poloidal magnetic fields, the underlying mechanism requires clarification. Fig.~\ref{fig:3}(b) reveals a diamagnetic cavity within the outflow, characterized by a depressed magnetic field relative to the 50 T background and surrounded by a region of enhanced magnetic pressure ($\text{B} > \text{B}_0$). The resulting magnetic pressure gradient exerts a confining $\text{J}\times \text{B}$ force, collimating the outflow, consistent with previous laboratory studies\cite{Lebedev+2005,Mabey+2019,Ivanov+2019,Malko+2024}. A critical aspect of this mechanism is the formation and evolution of the diamagnetic cavity itself.

\begin{figure}
\includegraphics[width=8cm]{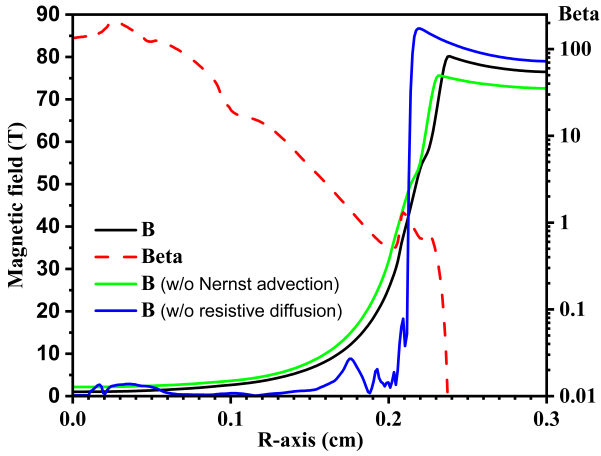}
\caption{Profiles of plasma beta ($\beta$) and magnetic field (B) across the radial direction at the half-height of the outflow for the $\text{B}_0$ = 50 T case at 30 ns. Results are compared for the regular simulation and cases excluding the Nernst effect or resistive diffusion.}
\label{fig:4}
\end{figure}

\textcolor{black}{Starting from the momentum equation of MHD equations (SI unit):
\begin{equation}
\rho \frac{d\mathbf{u}}{dt} = -\nabla p + \mathbf{J} \times \mathbf{B},
\end{equation}
where $\rho$ is the plasma mass density, $\mathbf{u}$ is the flow velocity, $\nabla \text{p}$ is plasma pressure gradient and $\mathbf{J}$ is plasma current density, then taking the cross product of both sides with $\mathbf{B}$, to get
\begin{equation}
\rho (\frac{d\mathbf{u}}{dt})\times \mathbf{B} = -(\nabla p)\times \mathbf{B} + (\mathbf{J} \cdot \mathbf{B})\mathbf{B}-B^2\mathbf{J},
\end{equation}
the second and third terms on right hand side can be simplified as $\text{J}_\parallel \text{B}\mathbf{B}-\text{B}^2(\mathbf{J_\parallel}+\mathbf{J_\perp})$, where $\mathbf{J_\parallel}$ and $\mathbf{J_\perp}$ are the currents parallel and perpendicular with magnetic field direction respectively. Therefore, the azimuthal current can be given by
\begin{equation}
\mathbf{J_\perp} = \frac{\mathbf{B} \times (\nabla p)}{B^2}  + \mathbf{B} \times \frac{\rho}{B^2} (\frac{d\mathbf{u}}{dt}),
\end{equation}
where on right hand side, the first term is the diamagnetic current ($\mathbf{J}_{\text{dia}}$)\cite{Chen+1984}, driven by the pressure gradient. The second term is the inertial current\cite{Heinemann+1994}, associated with plasma acceleration. Crucially, the magnetic field produced by these currents arises self-consistently from the MHD equations and is not a consequence of non-ideal induction terms. The role of diamagnetic currents is central to understanding collimation, as the quasi-equilibrium of solar magnetic flux tubes in the transverse direction is possibly sustained by surface diamagnetic currents.} 

\begin{figure}
\includegraphics[width=8cm]{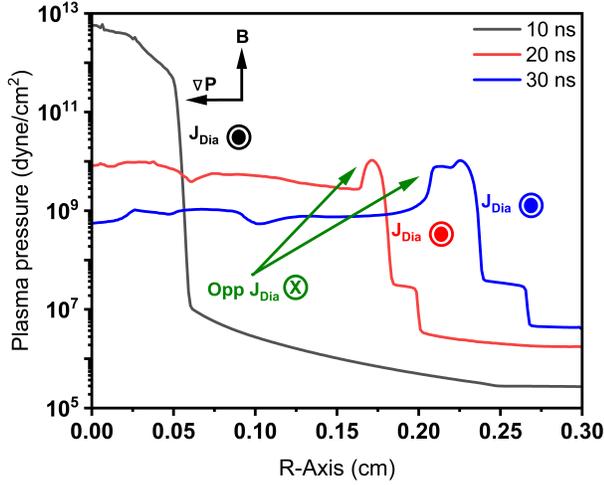}
\caption{\textcolor{black}{Plasma pressure profile across the radial direction at the mid-height of the outflow for the $\text{B}_0$ = 50 T caseat simulation times of 10, 20 and 30 ns. The $\mathbf{J}_{\text{dia}}$ indicate the diamagnetic current responsible for expelling the magnetic field. The opposing diamagnetic current (Opp $\mathbf{J}_{\text{dia}}$) arises from the pressure gradient directed toward the cavity shell.}}
\label{fig:5} 
\end{figure}

\begin{figure}
\includegraphics[width=8cm]{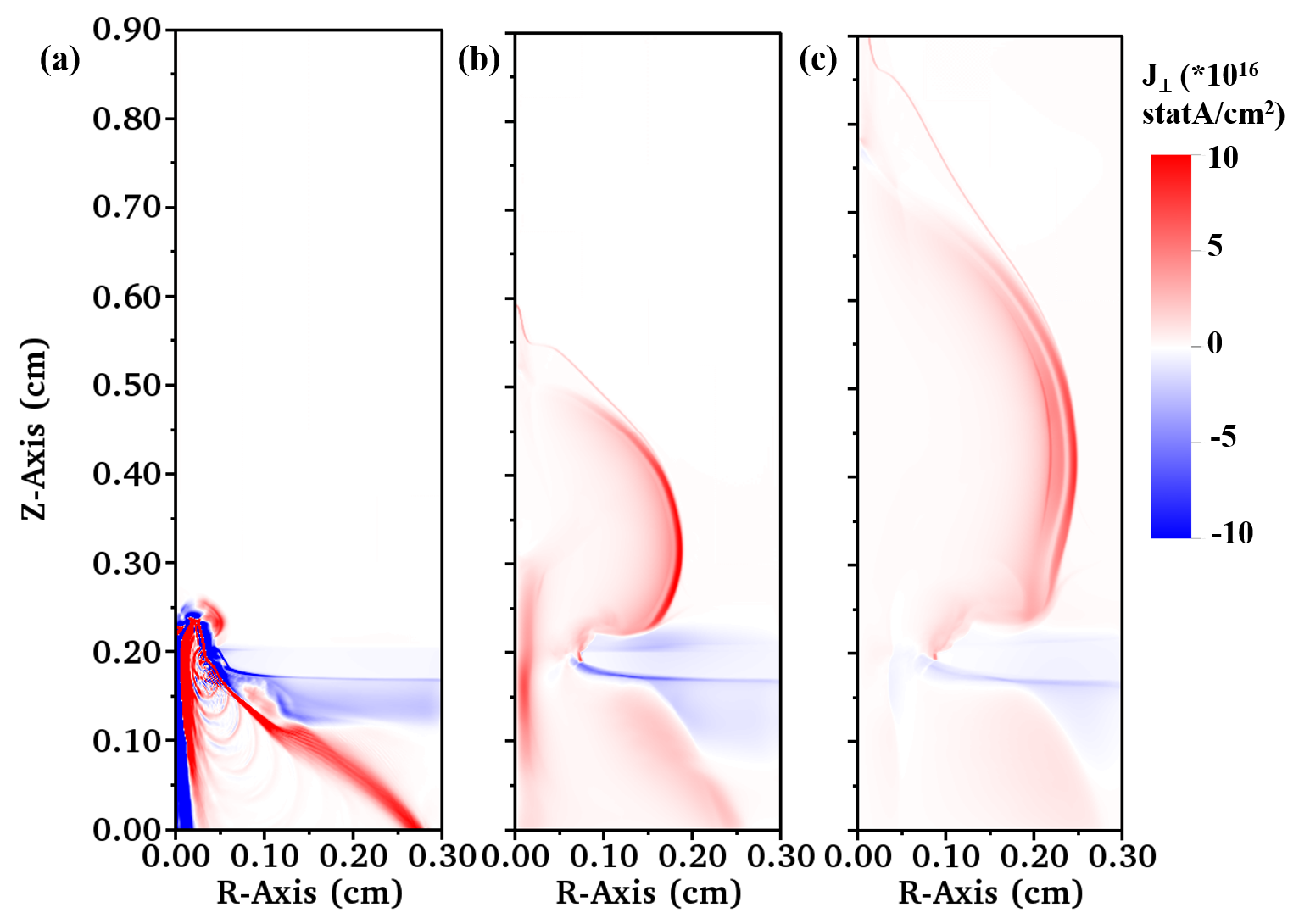}
\caption{\textcolor{black}{Azimuthal current density for the $\text{B}_0$ = 50 T case at (a) 10ns, (b) 20ns and (c) 30ns. Red regions correspond to current directed out of the plane, expelling magnetic field from the cavity. The outermost thin high-current layer is primarily composed by the diamagnetic current.}}
\label{fig:6} 
\end{figure}

\textcolor{black}{Simulations reveal that the cavity forms as the propagating outflow advects the magnetic field from its center to its periphery, a process further amplified and sustained by the azimuthal diamagnetic current, which is generated by the inward radial pressure gradient under the longitudinal magnetic field. This azimuthal diamagnetic current acts to expel the magnetic field from the outflow interior and amplify it externally, thereby reinforcing the confining magnetic pressure gradient. Specifically, during the laser heating period, the pressure of the initial plasma outflow decreases radially outward (Fig.~\ref{fig:5}, black curve). Thus, the plasma pressure gradient points entirely inward, causing the diamagnetic current to expel the magnetic field outward (Fig.~\ref{fig:6}a), with the excluded field accumulating outside. Subsequently, as the cavity's high-density shell forms due to confinement by the ambient magnetic field, a pressure gradient directed from the interior toward the shell develops (Fig.~\ref{fig:5}, red and blue curves), generating an opposing diamagnetic current. However, this opposing current does not offset the field-expelling current, allowing the mechanism to persist. The net azimuthal current involved in the $\text{J}\times \text{B}$ force remains outwardly directed, and the outermost thin high-current layer is primarily composed of diamagnetic currents (Fig.~\ref{fig:6}(b) and 6(c)).}

The plasma $\beta$ profile, shown in Fig.~\ref{fig:3}(c) and Fig.~\ref{fig:4}, drops rapidly from $\beta $>1 to $\beta $<1 in the radial direction. The transition from high to low $\beta$ coincides with the location where $\text{B} > \text{B}_0$, confirming that the confining region is a low-beta, magnetic-pressure-dominated environment due to plasma diamagnetism, \textcolor{black}{analogous to the transition} from photospheric to coronal plasma\cite{Gomez+2019}.

While the magnetic pressure gradient driven by diamagnetism is the primary collimation factor, non-ideal MHD effects can influence the magnetic field evolution~\cite{Sadler+2021}. The Biermann battery effect, describing the self-generated magnetic fields by misaligned density and temperature gradients, generates negligible fields (0.1 T at 30 ns). As shown in Fig.~\ref{fig:4}, the Nernst effect that \textcolor{black}{transports field in the opposite direction of the temperature gradient}, slightly increases the peak magnetic field at the cavity edge. \textcolor{black}{Resistive diffusion reduces boundary magnetic field by dissipation from ambient high magnetic pressure region to diamagnetic cavity center. However, it minimally affects the global field structure (Fig.~\ref{fig:4} blue curve). The magnetic Reynolds number ($\text{R}_\text{m}$) is given by
\begin{eqnarray}
R_m=uL/\eta , 
\end{eqnarray}
where L = 400 $\mu$m is the characteristic length determined by the thickness of the high density edge of jet, $\eta $ is the magnetic resistivity and $\text{ln}\Lambda $ is the Coulomb logarithm:
\begin{eqnarray}
\eta =&&8.22\times 10^{5}Zln\Lambda T_{e}^{-3/2}, \\
ln\Lambda =&&23.5-ln(N_{e}^{1/2}T_{e}^{-5/4})\nonumber\\
&&-\sqrt{10^{-5}+(ln(T_{e})-2)^2/16} , 
\end{eqnarray}
where Z is the average ionization and $\text{T}_\text{e}$ is the electron temperature. In the edge region of the diamagnetic cavity, $\text{R}_\text{m} > 10$ (Table~\ref{table:1}) indicating that advection dominates over diffusion, and the diffusion time scale ($\text{t}_\text{d}$)
\begin{eqnarray}
t_d=L^2/\eta 
\end{eqnarray}
is approximately $10^2$ ns. Given that the analysis focuses on 30 ns, the high magnetic field from edge does not have time to diffuse significantly into the cavity center or outward. It only slightly modifies the field gradient in the interior, without disrupting the global cavity structure.} \textcolor{black}{Therefore}, while these \textcolor{black}{non-ideal MHD} effects subtly alter the field distribution, they do not change the fundamental diamagnetic cavity structure or the overall collimation mechanism.

\begin{figure}
\includegraphics[width=8cm]{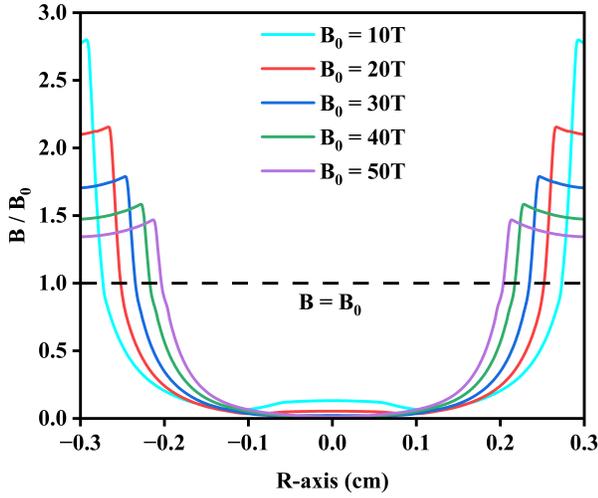}
\caption{Radial profile of the scaled magnetic field strength ($\text{B/B}_0$) at the outflow's half-height at 24 ns. The black dashed line indicates the position where $\text{B=B}_0$.}
\label{fig:7} 
\end{figure}

Diamagnetism is a robust feature observed across all simulated outflows, for applied fields ($\text{B}_0$) ranging from 10 to 50 T (Fig.~\ref{fig:7}). In all cases, the magnetic field is effectively expelled from the outflow center ($\text{B}/\text{B}_0 \approx 0$ for $\text{R} < 0.1 \text{cm}$), while the ambient field is amplified at the periphery ($\text{B}/\text{B}_0 > 1$). This confirms that collimation via diamagnetic cavity formation is a stable mechanism, largely independent of the specific $\text{B}_0$ strength. The high ambient magnetic pressure ($\text{B}/\text{B}_0 > 1$) only exist in a limited region around the diamagnetic cavity, and the $\text{B}/\text{B}_0$ will asymptote to 1 at large radial distances from the cavity. In addition, the cavity width is set by a pressure balance between the plasma kinetic pressure and the external magnetic pressure. As the laser energy (and thus plasma pressure) is constant, increasing $\text{B}_0$ from 10 to 50 T enhances the confining magnetic pressure, reducing the cavity width by approximately $25.7 \%$. This diamagnetic cavity physics is also important to other studies of laboratory and space plasma\cite{Winske+2019}.

\section{Discussion}\label{section5}
\textcolor{black}{Radiative cooling should be considered as a potential collimation mechanism when high-Z materials are used in laser targets, as it can reduce plasma pressure and enhance jet collimation\cite{Purvis+2010,Nicolai+2006,Yuan+2018,Gregory+2014}, thereby lowering the required external magnetic field strength\cite{Lei+2020}. In our simulations, however, radiative cooling is not significant contributed to the observed collimation, the radiation cooling time\cite{Ryutov+2011} is longer than 100 ns. This is because the simulation target is composed solely of CH plastic, excluding high-Z elements. Consequently, this work isolates and investigates the role of external magnetic pressure as the dominant collimation mechanism.}

\textcolor{black}{Building on the scaling theories of Ryutov et al\cite{Ryutov+2000,Ryutov+2001} and Falize et al\cite{Falize+2011}, Cross et al\cite{Cross+2014} further extended the framework to include quantum effects and systematically compared dimensionless parameters between laboratory and astrophysical plasmas.} To establish the astrophysical relevance of our \textcolor{black}{simulations}, key parameters of the laser-plasma outflow are compared with typical coronal outflows in Table~\ref{table:1}. \textcolor{black}{The coronal parameters, summarized from various observations\cite{Raouafi+2016,Shen+2021,Yang+2020}, correspond to the intended ejection region in the low-to-mid corona (< 3 $\text{R}_{\odot}$).} The Alfvén speed ($\text{V}_\text{A}$) \textcolor{black}{is calculated as
\begin{eqnarray}
V_A=2.18\times 10^{11}B/\sqrt {N_iA}, 
\end{eqnarray}
where $\text{N}_\text{i}$ is the ion number density, A is the average atomic weight. $\text{V}_\text{A}$ exhibits a sharp gradient at the cavity boundary due to jumps in density and magnetic field. The corresponding Alfvénic Mach number $\text{M}_\text{A}$,
\begin{eqnarray}
M_A=u/V_A,
\end{eqnarray}
indicates a radial transition from} super-Alfvénic \textcolor{black}{($\text{M}_\text{A}$ >1)} to sub-Alfvénic \textcolor{black}{($\text{M}_\text{A}$ <1).} Furthermore, the Reynolds number \textcolor{black}{($\text{R}_\text{e}$) is given by
\begin{eqnarray}
R_e=uL/\upsilon , 
\end{eqnarray}
where $T_i$ is the ion temperature, $\upsilon$ is the kinematic viscosity: 
\begin{eqnarray}
\upsilon =1.92\times 10^{19}\frac{T_{i}^{5/2}}{A^{1/2}Z^4N_{i}ln\Lambda }, 
\end{eqnarray}}
$\text{R}_\text{e}$ >> 1 in the simulation indicates that viscous dissipation is negligible, mirroring the conditions in coronal environments. The overall similarity in these \textcolor{black}{characteristic} parameters supports the scaling of our simulations. Consequently, the diamagnetic collimation mechanism demonstrated here provides a viable explanation for the formation of collimated outflows in the solar corona.

\begin{figure}
\includegraphics[width=4cm]{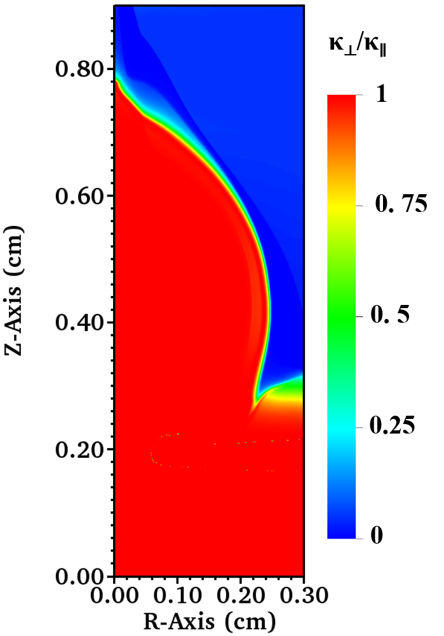}
\caption{\textcolor{black}{The ratio of perpendicular to parallel thermal conductivity $\kappa _\perp / \kappa _\parallel$ for the $\text{B}_0$ = 50 T case at 30 ns. $\kappa _\perp / \kappa _\parallel$ is close to 1 in most areas of outflow.}}
\label{fig:8} 
\end{figure}

\textcolor{black}{The isotropic thermal conduction model\cite{Atzeni+2004} is included in simulations. For the $\text{B}_0$ = 50 T case, the Hall parameter $\text{C}_\text{H}$,
\begin{equation}
C_H = 6.16\times 10^{16} \frac{T_{e}^{3/2}B}{Zln\Lambda N_{e}},
\end{equation}
varies from 0.1 in the cavity center to 2-3 at its edge. This suggests magnetic effects will be slightly displayed on thermal transport\cite{Sadler+2021}, where electron thermal conduction across field lines ($\kappa _\perp$) would be suppressed\cite{Matsuo+2017}. However, the ratio of perpendicular to parallel thermal conductivity $\kappa _\perp / \kappa _\parallel$\cite{Frank+2024} of the outflow is estimated to be 0.9 in most regions of the outflow and 0.3 only in the thin edge region, as shown in Fig.~\ref{fig:8}. The impact of suppressed $\kappa _\perp$ for the heat flux $\mathbf{q} = -\kappa _\parallel \mathbf{\hat{b}}(\mathbf{\hat{b}} \cdot \nabla \text{T}_\text{e}) - \kappa _\perp \mathbf{\hat{b}}\times (\nabla \text{T}_\text{e} \times \mathbf{\hat{b}})$ is not significant, in which $\mathbf{\hat{b}}$ is the unit vector in the direction of the magnetic field. Therefore, the usage of isotropic thermal conduction does not alter the fundamental conclusions regarding diamagnetic cavity formation and collimation.}

\textcolor{black}{The Peclet number $\text{P}_\text{e}$, as representing the ratio of thermal advection to thermal diffusion,is given by
\begin{equation}
P_e = Lu/\alpha,
\end{equation}
where the $\alpha$ is thermal diffusivity. $\text{P}_\text{e}$ is order of < 1 in our laboratory system thus the thermal diffusion is dominated, however, $\text{P}_\text{e}$ in solar corona is >> 1, indicating thermal advection dominates in the corona. This discrepancy in energy transport scaling is a recognized limitation when extending our results to the corona. Despite this, simulations achieve astrophysical relevance in key dynamic respects: the plasma temperature at the cavity edge (27–80 eV) is of the close order as in coronal outflows (120–138 eV). This ensures similarity in the sound speed $\text{C}_\text{S}$ and Mach number M
\begin{eqnarray}
C_S&=&9.79\times 10^5\sqrt{\gamma ZT_{e}/A}, \\
M&=&u/C_S,
\end{eqnarray}
where $\gamma$ is the adiabatic index, partially offsetting the transport limitation.} 

\begin{table}
\caption{Characteristic physical parameters of the laser-plasma outflow when $\text{B}_0$ = 50 T at 30 ns. The region inside the diamagnetic cavity ($\beta >1$) is excluded due to its low magnetic field strength, which does not meet the scaling requirements. Corresponding parameters for coronal outflows are summarized or calculated from observations\cite{Raouafi+2016,Shen+2021,Yang+2020}.}
\label{table:1}
\begin{ruledtabular}
\begin{tabular}{lcc}
\mbox{Plasma properties}&\mbox{Laser-plasma outflow}&\mbox{Coronal outflow}\\
\hline
\mbox{Average ionization}&\mbox{2-2.5}&\mbox{1}\\
\mbox{Average atomic weight}&\mbox{6.5}&\mbox{1.29}\\
\mbox{(a.m.u)}&\mbox{ }&\mbox{ }\\
\mbox{Number density $(1/cc)$}&\mbox{$0.1-3\times 10^{20}$}&\mbox{$0.9-2.8\times 10^8$}\\
\mbox{Temperature (eV)}&\mbox{27-80}&\mbox{120-138}\\
\mbox{Magnetic field (G)}&\mbox{$4-8\times 10^5$}&\mbox{1-4}\\
\mbox{Beta}&\mbox{<1}&\mbox{<1}\\
\mbox{Flow speed (km/s)}&\mbox{40-300}&\mbox{200-250}\\
\mbox{Sound speed (km/s)}&\mbox{40-150}&\mbox{200}\\
\mbox{Alfvén speed (km/s)}&\mbox{50-1700}&\mbox{1000}\\
\mbox{Alfvénic Mach number}&\mbox{0.02-3}&\mbox{0.4-0.5}\\
\mbox{Mach number}&\mbox{2-5}&\mbox{1-1.25}\\
\mbox{Reynolds number}&\mbox{$2-11\times 10^{5}$}&\mbox{$1.7\times 10^3$}\\
\mbox{Magnetic Reynolds number}&\mbox{27-180}&\mbox{$3.5\times 10^{12}$}\\
\end{tabular}
\end{ruledtabular}
\end{table}

\textcolor{black}{The diamagnetic cavity is conceptually divided into two distinct radial regions: (1) the central core, characterized by a strongly suppressed magnetic field (a few Tesla) and a high plasma-beta (tens to hundreds); and (2) the cavity edge, where the magnetic field recovers to near or exceeds the background level and beta is on the order of or below unity. Data from the high-beta central core are intentionally excluded from the scaling analysis because this region does not satisfy the fundamental coronal scaling condition of magnetic pressure dominance. However, this exclusion in no way diminishes the astrophysical relevance of the diamagnetic cavity model. The formation of a low-magnetic-field cavity core is an intrinsic consequence of plasma diamagnetism. The prevalence of diamagnetic structures in the corona is confirmed by observational evidence such as the "melon-seed" effect\cite{Pneuman+1984,Panasenco+2019} and diamagnetic currents on the coronal flux tubes\cite{Zaitsev+2005,Parkhomov+2018}. Crucially, the laboratory condition where the diamagnetic cavity is radially confined by external magnetic pressure is directly analogous to the proposed situation for coronal outflows\cite{Moore+2013}. This parallel reinforces the physical connection between our scaled simulation and the coronal outflow.}

\section{Summary}\label{section6}
The collimation mechanism of coronal outflows remains an open question. While ambient magnetic pressure is a proposed key factor for coronal jets, it requires \textcolor{black}{discussion} through scaled laboratory \textcolor{black}{studies. This numerical study pursues four primary objectives: First, to produce a stable, persistent, and readily diagnosable diamagnetic cavity under a long pulse laser facility (like Omega) condition. Second, to reproduce (scaled) plasma conditions that approximate those of solar coronal outflows, particularly in Alfvén speed and beta. Third, to elucidate the formation of the diamagnetic cavity and the concomitant diamagnetic amplification of the ambient magnetic field, thereby clarifying the magnetic-pressure-driven collimation mechanism. Finally, this work uses FLASH simulations to plan future experiments on laser facilities.}

Simulations of a laser-driven plasma \textcolor{black}{were completed by FLASH code} to model a coronal outflow following open magnetic field lines. \textcolor{black}{Simulation} results demonstrate that the outflow is collimated by a diamagnetic cavity structure, where magnetic field is advected from the center to the edge. Diamagnetic currents further depress the internal field and amplify the external field, creating a strong magnetic pressure gradient that confines the flow. This mechanism is robust across applied fields up to 50 T, with stronger fields yielding tighter collimation and a smaller cavity radius—consistent with the balance requirement between plasma pressure and external magnetic pressure. Non-ideal MHD effects such as Biermann battery, resistive diffusion and Nernst effect are found to be negligible. \textcolor{black}{The simulation is scaled to the corona by preserving key dimensionless parameters, including the Alfvén speed and plasma beta.} The similarity in plasma conditions confirms that the ambient magnetic-pressure-dominated, low-beta plasma condition is responsible for the collimation of the coronal outflow, as demonstrated by \textcolor{black}{the} mechanism \textcolor{black}{displayed in this work.} Future research can further explore the impact of more complex coronal magnetic field geometries on the outflow morphology, such as bent field lines, coronal loops, effects of turbulence and low collisionality.

\begin{acknowledgments}
This work was supported by the Delaware Space Grant Consortium 80NSSC19M0087 and the Bartol Research Institute. The software used in this work was developed in part by the DOE NNSA- and DOE Office of Science-supported Flash Center for Computational Science at the University of Chicago and the University of Rochester. \textcolor{black}{The authors thank the anonymous referees for their insightful comments that helped improve the manuscript.}
\end{acknowledgments}

\section*{Data Availability Statement}
The data that support the findings of this study are available from the corresponding author upon reasonable request.

\nocite{*}

\bibliography{Reference}

\end{document}